%% file: CC-model-1-aps.tex
\documentclass[prl,twocolumn,showpacs,superscriptaddress]{revtex4-1}
\usepackage{natbib}
\usepackage[english]{babel}
\usepackage{amsfonts,amsmath,bm}
\usepackage[fleqn,tbtags]{mathtools}
\usepackage{graphicx}
\usepackage[usenames,dvipsnames]{xcolor}
\usepackage[utf8]{inputenc}

\usepackage{tikz,pgfplots}
\usetikzlibrary{matrix,arrows}

\usepackage{hyperref}

\newcommand{\be}{\begin{equation}}
\newcommand{\ee}{\end{equation}}
\newcommand{\bes}{\begin{equation}\begin{split}}
\newcommand{\ees}{\end{split}\end{equation}}
\newcommand{\bea}{\begin{eqnarray}}
\newcommand{\eea}{\end{eqnarray}}

\newcommand{\rez}[1]{\frac{1}{#1}}

\def\beq{\begin{equation}}
\def\eeq{\end{equation}}
\def\bea{\begin{eqnarray}}
\def\eea{\end{eqnarray}}

\begin{document}

\title{Localization length index
in a Chalker-Coddington model: a numerical study}

\author{W. Nuding}

\affiliation{Wuppertal University, Gaußstraße 20, Germany}

\author{A. Klümper}

\affiliation{Wuppertal University, Gaußstraße 20, Germany}

\author{A. Sedrakyan}

\affiliation{Wuppertal University, Gaußstraße 20, Germany}

\affiliation{Yerevan Physics Institute, Br. Alikhanian 2, Yerevan 36, Armenia}

\begin{abstract}
	We calculated numerically the localization length index $\nu$ for
	the Chalker-Coddington model of the plateau-plateau transitions 
	in the quantum
	Hall effect. By taking into account finite size effects we have obtained
	$\nu = 2.593 \pm 0.0297$. The calculations were
	carried out by two different programs that produced close results, each one
	within the error bars of the other. We also checked the possibility of
	logarithmic corrections to finite size effects and found, that they come
	with much larger error bars for $\nu$.
\end{abstract}


\maketitle

The computation of critical indices of the plateau-plateau transitions in the
quantum Hall effect (QHE) (see for a review \cite{huckestein1995}) is still an
open problem in modern condensed matter physics.  According to the pioneering
works on localization \cite{anderson1958} the dimension two is a marginal
dimension, above which delocalization can appear.  Exactly at d=2 Levine,
Libby and Pruisken
\cite{Levine1984yg,Levine1984yf,Levine1984ye} noticed, that the presence of a
topological term in the nonlinear sigma model (NLSM) formulation of the
problem may result in the appearance of delocalized states in strong magnetic
fields.  The next achievement was reached
by Chalker and Coddington \cite{chalker1988percolation}. 

The authors formulated and studied numerically a network model ($CC$ model) in
a random potential yielding localization-delocalization transitions. The
numerical value $2.5\pm 0.5$ of the Lyapunov exponent (LE) in the CC model was
in good agreement with the experimentally measured localization length index
$\nu= 2.4$ in the quantum Hall effect \cite{wei1994current}.  Recently the
more precise value $ \nu =2.38 \pm 0.06$ was reported in \cite{li2005scaling,
  li2009scaling}.

Various aspects of the CC-model were investigated in a chain of interesting
papers: In \cite{lee1994network} the model was linked to replicated
spin-chains, while in \cite{zirnbauer1994jg,zirnbauer1997} its connection to
supersymmetric spin-chains was revealed. Some links with conformal field
theories of Wess-Zumino-Witten-Novikov (WZWN) type were presented in
\cite{tsvelik2007} and \cite{leclair2001}.

In Refs.\ \cite{obuse2012,evers2008multifractality} the authors investigated
the multifractal behaviour of the CC model. Both papers reported quartic
deviations from the exact quadratic dependence of the multifractal indices on
the parameter $q$, which was predicted in
Refs.\ \cite{tsvelik2007,leclair2001}. This fact points out that the validity
of the simple, supersymmetric WZWN approach to plateau-plateau transitions in
the quantum Hall effect is questionable and here we are still far from the
application of conformal field theory.

In spite of a lot of understanding that has been gained for the
plateau-plateau transitions in the QHE, the final model which would allow for
the calculation of the localization length index either analytically or
numerically has not been formulated yet. Moreover, recently more precise
numerical calculations of the localization length index of the CC-model
\cite{slevin2009, amado2011numerical, obuse2012, dahlhaus2011} show values
close to $2.61 \pm 0.014$, which is well far from the experimental value $2.38
\pm 0.06$ \cite{li2009scaling}.  This indicates that the CC-model as such is
not applicable to the description of plateau transitions.

\begin{figure}[t]
	\centerline{\includegraphics{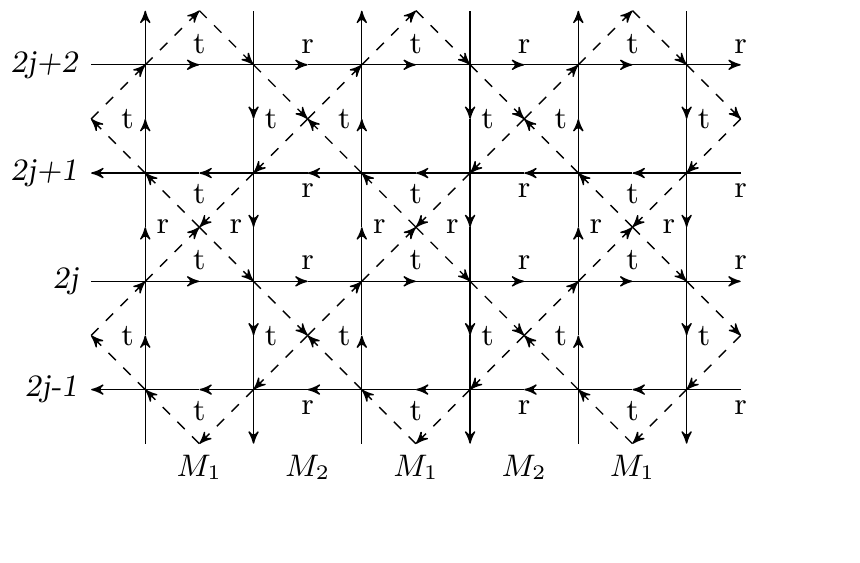}}
	\caption{Schematic illustration of the CC network. $M_1$ and $M_2$ denote
		the column transfer matrices as defined in \eqref{M1} and \eqref{M2}. Multiplication
		with a column transfer matrix describes the transition of a particle through the corresponding
		column of the lattice.}
	\label{figML}
\end{figure}

Up to now all numerical analyses of finite size scaling in the CC-model
\cite{slevin2009, amado2011numerical, obuse2012, dahlhaus2011} show that the
second, irrelevant operator in the model has a scaling dimension very close to
the major one.  Moreover, in ~\cite{amado2011numerical} it was claimed that
the next to leading order finite size resp.~width $M$ corrections have
$1/\log[M]$-form, which indicates for the CC-model the possible presence of
two operators with almost equal conformal dimensions.

The goal of the current paper is threefold: First we want to recalculate the
localization length index in order to test the results obtained in
\cite{slevin2009, amado2011numerical, obuse2012, dahlhaus2011}. Second we
want to check whether the $1/\log[M]$-form for the corrections is adequate
or not. Third we want to explore the possibility of two irrelevant fields
in the scaling analysis. 

To 
achieve these goals, we developed two
independent codes to numerically investigate the finite size scaling of the
CC-model. We calculated both the localization length index and the next to
leading index.

For the calculation of critical indices we used the transfer-matrix method
developed in \cite{mackinnon1981scaling, mackinnon1983scaling}.  We had to
calculate the smallest Lyapunov exponent (LE) of the CC-model, for which it was
necessary to calculate a product $T_L=\prod_{j=1}^L M_1 U_{1j}M_2 U_{2j}$ of
layers of transfer matrices $M_1 U_{1j}M_2 U_{2j}$ corresponding to two
columns $M_1$, $M_2$ of vertical sequences of 2x2 scattering nodes,
cp.~Fig.~\ref{figML}:
\begin{equation} \label{M1}
M_1= \begin{tikzpicture}[baseline=(current bounding box.center), ultra thick, loosely dotted]
				\matrix (M1) [matrix of math nodes,nodes in empty cells,right delimiter={)},left delimiter={(}] {
					B^1 & 0   & 	& 0  	\\
					0   &	B^1	&  	& 		\\
							& 		& 	& 0 	\\
					0 	& 		& 0	& B^1	\\
				};
				\draw (M1-2-2)--(M1-4-4);
				\draw (M1-1-2)--(M1-1-4);
				\draw (M1-4-1)--(M1-4-3);
				\draw (M1-2-1)--(M1-4-1);
				\draw (M1-1-4)--(M1-3-4);
				\draw (M1-1-2)--(M1-3-4);
				\draw (M1-2-1)--(M1-4-3);
			\end{tikzpicture}
\end{equation}
and
\begin{equation}\label{M2}
	M_2=\begin{tikzpicture}[baseline=(current bounding box.center), ultra thick, loosely dotted]
				\matrix (M2) [matrix of math nodes,nodes in empty cells,right delimiter={)},left delimiter={(}] {
					B^2_{22}	&	0  	&		&	0	&	B^2_{21}		\\
					0					&	B^2	&		&			&	0					\\
					 					& 		&		&			&						\\
					0      		&			&		&	B^2	&	0					\\
					B^2_{12}	&	0		&		& 0		&	B^2_{11}	\\
				};
				\draw (M2-1-2)--(M2-1-4);
				\draw (M2-1-2)--(M2-4-5);
				\draw (M2-2-1)--(M2-4-1);
				\draw (M2-2-1)--(M2-5-4);
				\draw (M2-2-2)--(M2-4-4);
				\draw (M2-2-5)--(M2-4-5);
				\draw (M2-5-2)--(M2-5-4);
			\end{tikzpicture}
\end{equation}
with
\begin{equation}
	B^1=\begin{pmatrix}
				1/t & r/t \\
				r/t & 1/t 
			\end{pmatrix}
	\qquad \text{and} \qquad 
	B^2=\begin{pmatrix}
				1/r & t/r \\
 				t/r & 1/r
			\end{pmatrix}
\end{equation}
where periodic boundary conditions were imposed on $M_2$. The $U$-matrices have
a simple diagonal form:
$\left[U_{1,2}\right]_{nm}=\exp{(i\alpha_n)}\,\delta_{nm}$.  Here $t$ and $r$
are the transmission and reflection amplitudes at each node of the regular
lattice shown in Fig.~\ref{figML} which are suitably parameterized by
\begin{equation} \label{rt}
  t=\rez{\sqrt{1+e^{2x}}} \qquad \text{and} \qquad r=\rez{\sqrt{1+e^{-2x}}}.
\end{equation}
The model parameter $x$ corresponds to the Fermi energy measured from the
Landau band center scaled by the Landau band width (so the critical point is
$x=0$) while the phases $\alpha_{n}$ are stochastic variables in the range
$[0,2\pi)$, reflecting the randomness of the smooth electrostatic potential
landscape.

We calculated the product of a chain of transfer matrices which contain random
parameters. According to the Oceledec
theorem~\cite{oseledec1968multiplicative} the $\frac{1}{L}$ power of the
product has a set of eigenvalues, which are independent of the history of the
randomness.  The logarithms of the moduli of these eigenvalues are called
Lyapunov exponents.  \bea
\label{LE}
\gamma=\lim_{L\to\infty}\frac{\log[T_L T_L^\dagger]}{2L}\ ,
\eea

The smallest positive one of these exponents yields
the critical behaviour of the correlation length
of the model, i.e. $\gamma \sim x^{-\nu} $ where $\nu$
is the localization length index.

It is clear, that numerically the infinite limit cannot be calculated.  For
chains with finite length $L$, the central limit
theorem~\cite{tutubalin1965limit} tells us that the Lyapunov exponents have a
Gaussian distribution with variance $\sigma_{\gamma} \sim \sqrt{\frac{M}{L}}$.

This means, that by considering a chain of length $L$ we calculate the LE with
error $\sim \sqrt{\frac{M}{L}}$.  Moreover, if we consider an ensemble of $N$
chains, the variance becomes $\sim \sqrt{\frac{M}{L N}} $.  Therefore our
strategy will be to consider large ensembles of chains.

We used ensembles of products with length $L$ ranging from $1\,000\,000$ to
$5\,000\,000$. The details about our data base can be found in table
\ref{statistics}.

\begin{table}
\input{table_statistics}
\caption{This table shows the statistics of the data. For each
    $M$ we have calculated the Lyapunov exponent with the 13 $x$-values that
    divide the interval $[0,0.08]$ into 12 equal parts.}
 \label{statistics}
\end{table}

Calculating these matrix products the naive way is not possible as
many entries of the product very soon exceed the size of all available data
types.  One can overcome this problem by use of the
method presented in \cite{mackinnon1981scaling, mackinnon1983scaling},
namely, the product can be performed with repeated QR decompositions. The
rightmost $T$ is QR decomposed. The unitary Q is then multiplied with the
next $T$ and the product is decomposed again. Repeating this procedure many
times we are in principle left with some Q multiplied with a product of
upper right triangular matrices.  It appears, that
the product of the diagonal entries of the upper
triangular matrices are approaching the eigenvalues of the total transfer
matrix $T_L$.
Of these numbers we are only interested in those which are close to 1.

For details see for instance \cite{vonBremen1997}. In our numerical 
simulations  we found that it is also possible 
to apply the much faster LU decomposition instead of the QR decomposition.

\section{The fitting procedure}
From the scaling behaviour of the Lyapunov exponent near the critical point 
we expect for finite size systems
\begin{equation} \label{ren_equ}
	\gamma M=\Gamma(M^{1/\nu}u_0,f(M)\,u_1) ,
\end{equation}
where $f(M)$ is decreasing with $M$. Here $M$ is the
number of nodes in each column of the lattice. $u_0=u_0(x)$ is a relevant
field and $u_1=u_1(x)$ the leading irrelevant field. It is common to choose
$f(M)=M^y$, $y<0$. Further it is known, that the relevant field vanishes at
the critical point.  The left hand side was obtained from \eqref{LE} dependent
on the parametrization parameter $x$ and the lattice height $M$. The right
hand side was expanded in a series in $x$ and $M$ and the coefficients were
obtained by a fit. Some coefficients in this expansion need not to be taken
into account as can be seen following the arguments of \cite{slevin2009}:

If $x$ is replaced by $-x$ we see from \eqref{rt} that $t$ turns into $r$ and
vice versa.  Due to the periodic boundary conditions the lattice is
unchanged. Therefore the left hand side of \eqref{ren_equ} is invariant under
a sign flip of $x$. Hence the right hand side must be even in $x$.  That makes
$u_0(x)$ and $u_1(x)$ even or odd. For the Chalker Coddington network the
critical point is at $x=0$. This makes us choose $u_o(x)$ odd and $u_1(x)$
even. The fit now should use as few coefficients as possible while reproducing
the data as good as possible.

One reasonable attempt is to do an expansion of the right hand side of
\eqref{ren_equ} in $x$.  This yields
\begin{equation}\label{G_expansion_x}
\begin{split}
	\Gamma=&\;\Gamma_{00}+\sum_{k=1}^\infty \Gamma_{0k}\,M^{ky} \\
	& +x^2\left[b_2\sum_{k=1}^\infty\Gamma_{01}\,M^{ky} + M^{2/\nu}\sum_{k=0}^\infty\Gamma_{2k}M^{ky} \right] \\
	& + x^4 \left[ (b_4+b_2^2)\sum_{k=1}^\infty \Gamma_{0k}M^{ky} +b_2M^{2/\nu}\sum_{k=1}^\infty 
		\Gamma_{2k}M^{ky}\right. \\
	& \qquad \ + \left. M^{4/\nu}\sum_{k=0}^\infty \Gamma_{4k}M^{ky} + a_3M^{2/\nu}\sum_{k=0}\Gamma_{2k}M^{ky} 
	  \right] \\
	& +O(x^6)
\end{split}
\end{equation}
A subset of this is the fitting formula used in \cite{obuse2012}. We tried
both formulas and the one in \eqref{G_expansion_x} worked out better for our
data. \\
The fitting formula above was derived by first expanding $\Gamma$ in the
fields
\begin{equation}
\label{expansin_in_fields}
\begin{split}
	\Gamma(u_0(x)&M^{1/nu},u_1(x)M^y)= \Gamma_{00}+ \Gamma_{01} u_1M^y \\
	& +\Gamma_{20}u_0^2M^{2/\nu}+	\Gamma_{02}u_1^2M^{2y} \\
	& +\Gamma_{21}u_0^2u_1M^{2/\nu}M^y +\Gamma_{03}u_1^3 M^{2y} \\
	& +\Gamma_{40}u_0^4M^{4/\nu}+\Gamma_{22}u_0^2u_1^2 M^{2y} + \Gamma_{04}u_1^4M^{4y} \\
	& +\dots
\end{split}
\end{equation}
and then the fields in $x$ like it has been done by most other authors
in the past
\begin{equation}
\label{fields_expanded}
 u_o(x)=x+\sum_{k=1}^\infty a_{2k+1}x^{2k+1} \quad \text{and} \quad
 u_1(x)=1+\sum_{k=1}^\infty b_{2k}x^{2k}
\end{equation}
In \eqref{expansin_in_fields} all coefficients in the expansion of $\Gamma$
that would contradict the scaling function being even in $x$ have been
dropped. Because of ambiguity in the overall scaling of the fields, the
leading coefficient in \eqref{fields_expanded} can be chosen to be 1. \\
Of course the described expansion is unique, however when taking into account
a finite number of expansion coefficients $\Gamma_{lk}$ and $a_n$, $b_m$,
different fitting procedures can be devised.  With  formula
\eqref{G_expansion_x} we obtained the best fits for our data.

We also considered the case of two irrelevant fields. This, in analogy to 
\eqref{ren_equ}, gives
\begin{equation}
	{\gamma M=\Gamma(M^{1/\nu}u_0,M^{y_1}\,u_1,M^{y_2}\,u_2), \quad y_1,y_2<0}
\end{equation}
With the same reasoning as in the case of one irrelevant field we find that $\Gamma$
is even in $x$. Along the lines of the above case we obtain that
$u_0$ is odd and $u_1$ and $u_2$ are even in $x$. Of course $\Gamma$ is even
in $x$, too. This helps to identify expansion coefficients that are zero like 
in the case of one irrelevant field.

\section{Results}

\begin{figure}[t]
	\centerline{\includegraphics{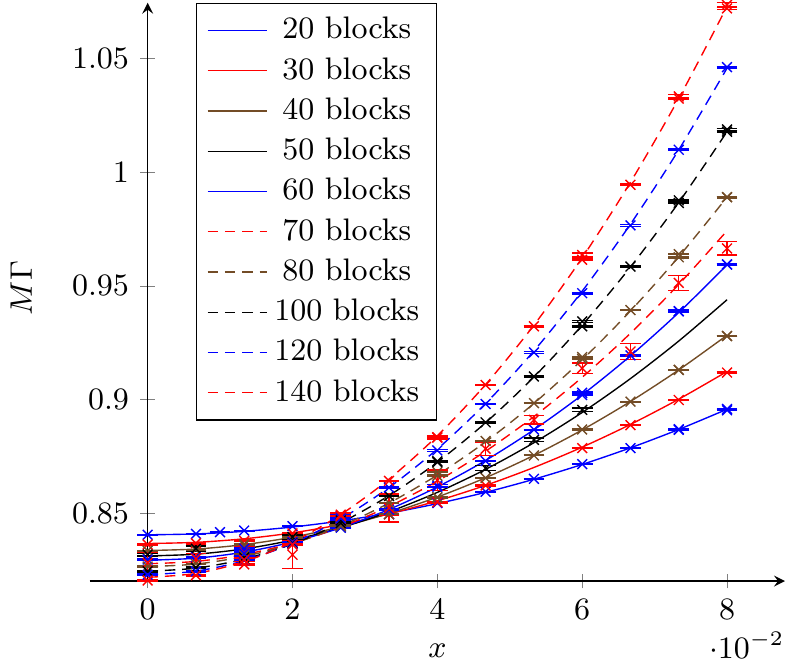}}
	\caption{Plot of the smallest eigenvalue of the transfer matrix for different
  	block sizes and in dependence on the distance $x$ from the critical
  	point. 
		The $x$-values divide the interval $[0,0.08]$ into 12	equal parts.}
	\label{fig2}
\end{figure}

In Fig.\ref{fig2} we present the leading Lyapunov exponent for various numbers
of $2 \times 2$ blocks in the transfer matrix versus $x$ (defined by formulas 
\eqref{rt}), which measures the deviation of the hopping parameters $r$ and $t$ 
from their critical value $1/\sqrt{2}$. The corresponding fitting parameters are 
presented in the table below.

\begin{figure}
	\centerline{\includegraphics{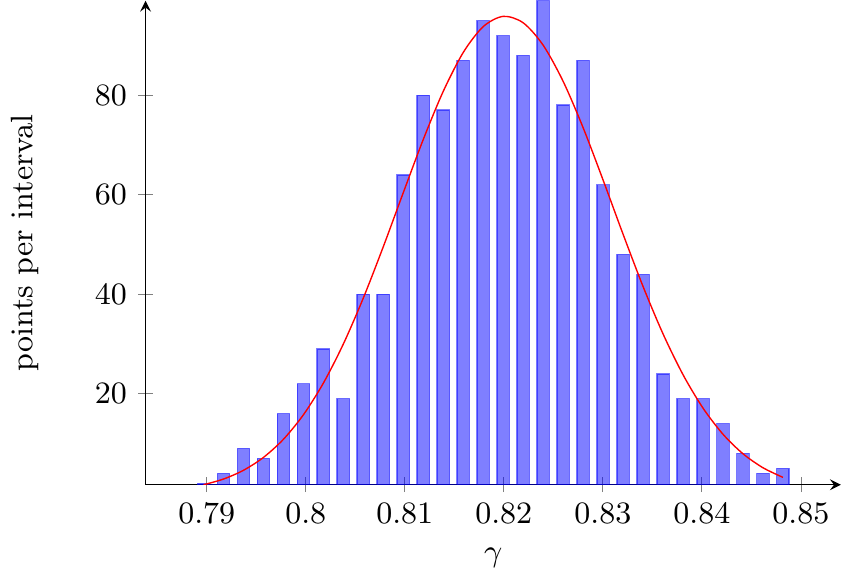}}
	\caption{Distribution of Lyapunov exponents in the ensemble of calculations 
		with 1282 elements for chain length $L=1\,000\,000$, $M=140$ and $x=0$}
	\label{fig1}
\end{figure}

In Fig.\ref{fig1} we present an example of the distribution
of Lyapunov exponents with fixed $M$, product length $L$ and $x$. This
distribution defines one point and its error for the fit. Here, we see a
Gaussian distribution in full accordance with the central limit theorem
\cite{tutubalin1965limit}.

The fits have been performed with a trust region algorithm. In a first step
the region for each parameter is chosen. Initial values within these regions
are taken at random. The results are the initial values for the next fit
without regional restrictions. These results are taken again as initial
values. This is done recursively 200 times.

Our best fitting results have been obtained by taking the first two lines of
\eqref{expansin_in_fields} and expanding $u_0$ up to the third and $u_1$ up to
the fourth order in $x$. 


\smallskip \noindent
For the fitting formula\,:
\vspace{-1ex}
\begin{equation}
	\begin{split}
		M\Gamma(x,M) =\; &\Gamma_{00}+\Gamma_{01}*(1+b_2*x^2)*(M^y) \\
		& +\Gamma_{20}*(x*M^{1/\nu})^2 \\
		& +\Gamma_{02}*((1+b_2*x^2)*(M^y))^2
	\end{split}
\end{equation}
we found the following coefficients and goodness of fit parameters\,: \\

\noindent
Coefficients (confidence bounds 95\%)\,:
\begin{align*}
			 \hline \\[-2.5ex]
       \Gamma_{00} =\; &   \quad   0.737 & (0.494&, 0.981) \\
       \Gamma_{01} =\; &   \quad  0.185 & (0.140&, 0.230) \\
       \Gamma_{02} =\; &   \quad  -0.0452 & (-0.300&, 0.209)  \\
       \Gamma_{20} =\; &   \quad  0.863 & (0.821&, 0.905) \\
       b_2 =\;&    -0.784 & (-5.533&, 3.964) \\
       \nu =\; &    \quad    2.59 & (2.560&, 2.619) \\
       y =\;&     -0.134 & (-0.607&, 0.339) \\
       \hline
\end{align*}
\noindent
Goodness of fit parameters\,:
\nopagebreak \vspace{-0.2ex}
\begin{align*}
  \hline \\[-2.5ex]
	&\text{sum of squares due to error\,: } && 2.03664\cdot 10^{-7} \\
  &\text{R-square\,: } && 0.999925 \\
  &\text{degrees of freedom adjusted R-square\,: } && 0.999924 \\
  &\text{root mean squared error\,: } && 1.85167\cdot10^{-5} \\
	&\text{sum of residuals\,: }&& 0.00297 \\
  &\text{degrees of freedom\,: }&& 594 \\
  \hline
\end{align*}

It turned out that the fit result depends slightly on the randomly chosen
initial values. That means the parameters turn out different if we fit the
same data several times. To take this into account we averaged over
200 fits as described above. Of course the averaged
set of coefficients is not a good parameter set for the fit as all
coefficients are highly correlated. So we just took the average for the
critical index $\nu$. The distribution of $\nu$-values showed to have a
Gaussian distribution.  The average for $\nu$ gave\,:
\begin{equation}
	\nu = 2.593     \pm 0.0297
\end{equation}
Here the error is given by the standard deviation of the sample for the $\nu$-values. This result is 
in perfect agreement with the other recent work like \cite{amado2011numerical,obuse2012,slevin2009}.
For $y$ we obtained analogously
\begin{equation}
  y = -0.145     \pm 0.0677
\end{equation}

We have also found, that the fit with $1/\log(M)$ corrections does not give
acceptably small confidence bounds for the main fitting value, the
localization length index $\nu$. All attempts with different numbers of
fitting coefficients did not lead to narrower confidence bounds.

Also potentially interesting is the ansatz with two irrelevant fields. It
reproduces $\nu=2.6$ quite well, too.  Averaging over an ensemble of fits
similar to the case of one irrelevant field yields
\begin{eqnarray}
	\nu & = & 2.608  \pm 0.0257 \\
	y_1 & = & -0.728 \pm 0.077  \\
	y_2 & = & -0.733 \pm 0.093  
\end{eqnarray}
Again the error is given by the standard deviation of the ensemble of
fits. $y_1$ and $y_2$ are quite similar 
in magnitude. We can neither explain
why this is the case nor do we have a theoretical reason for the presence of
two irrelevant fields. Identifying $y_1$ and $y_2$ 
is not equivalent with the 
case of a fit with one irrelevant field.

\section{conclusion and outlook}
Our main result is in perfect agreement with the values of
the localization length presented in the recent works \cite{slevin2009,
amado2011numerical, obuse2012, dahlhaus2011}.   

We have also tested the goodness of the fit with $1/\log(M)$ corrections and 
found, that a power behaviour of the second, sub-leading term in the Lyapunov 
exponent is preferable, though it is very small, $y \sim -0.145$, indicating 
its proximity to $f(M)= 1/\log(M)$ (see \eqref{ren_equ}) in the 
case of one irrelevant field.

We also successfully fitted two irrelevant fields. For this fit the confidence
bounds are much wider but the result for $\nu$
is less affected by different choices for the range of values for $M$. As a
fit with fewer coefficients is preferable we think that the ansatz with one
irrelevant field is better. But we cannot rule out the possibility that indeed
two irrelevant fields are important.
It is clear that new and
more extensive computations are needed to collect enough
statistics for distinguishing the indices of the irrelevant operators with
necessary precision.
 
The result confirms the necessity of an essential modification of the CC-model 
for the description of the plateau-plateau transition in the QHE.  

\subsection*{Acknowledgments}
\begin{acknowledgments}
A.~S.\ thanks the Theoretical Physics group at Wuppertal
University for the hospitality extended to him. A.~S.~and A.~K.~ acknowledge
support by DFG grant KL 645/7-1. The work of A.~S.\ was partially supported
by ARC grant 13-1C132. 
Most of the code development for cluster computation has been performed on the
Grid Cluster of the High Energy Physics group of the Bergische Universität
Wuppertal, financed by the Helmholtz - Alliance 'Physics at the Terascale' and
the BMBF. Computations have been carried out on Kaon (Wuppertal). Extensive
calculations have been performed on Rzcluster (Aachen), PC\textsuperscript{2} (Paderborn) and particularly on JUROPA 
(J\"ulich). The authors gratefully acknowledge the computing time 
granted by the John von Neumann Institute for Computing (NIC) and provided on the 
supercomputer JUROPA at Jülich Supercomputing Centre (JSC).
\end{acknowledgments}

\nocite{Levine1984yg,Levine1984yf,Levine1984ye}
\vfill
\bibliography{Literaturverzeichnis} 
\bibliographystyle{apsrev4-1}

\end{document}

%% file: table_statistics.tex
\begin{center}
\begin{tabular}{c|c|c|c}
	$M$ & $L$ & number of products & program \\
	\hline
	20	& 1\,000\,000 & 900		& Fortran		\\
	20  & 5\,000\,000 & 100   & C++      	\\
	40  & 1\,000\,000 & 1000  & Fortran   \\
	40  & 5\,000\,000 & 350		& C++				\\
	60	& 1\,000\,000 & 1000	& Fortran		\\
	60  & 5\,000\,000 & 280		& C++				\\
	80	& 1\,000\,000 & 1000	& Fortran		\\
	80	& 5\,000\,000 & 380		& C++				\\
	100	& 1\,000\,000 & 1020	& Fortran		\\
	100 & 5\,000\,000 & 150		& C++ 			\\
	120 & 1\,000\,000 & 850		& Fortran   \\
	120 & 5\,000\,000 & 300   & C++				\\
	140 & 1\,000\,000 & 1260  & Fortran   \\
	140 & 5\,000\,000 & 310		& C++				\\
	160 & 1\,000\,000 & 285   & Fortran   \\
	160 & 5\,000\,000 & 220 	& C++				\\
	180 & 1\,000\,000 & 240		& Fortran		\\
\end{tabular}
\end{center}